\documentclass[prl,english,superscriptaddress,twocolumn]{revtex4-1}

\usepackage{amsmath,amsfonts,amssymb}
\usepackage{graphicx,float,calc}
\usepackage[breaklinks,colorlinks,urlcolor=blue,citecolor=blue,linkcolor=blue]{hyperref}

\begin{document}

 \title{Exploring Parity Magnetic Effects through Experimental Simulation with Superconducting Qubits
 }
	
	\author{Yu Zhang}\thanks{These authors contributed equally to this work.}
	\affiliation{National Laboratory of Solid State Microstructures, School of Physics,
		Nanjing University, Nanjing 210093, China}
	
	\affiliation{Shishan Laboratory, Suzhou Campus of Nanjing University, Suzhou 215000, China}
	
	\author{Yan-Qing Zhu}\thanks{These authors contributed equally to this work.}
\affiliation{Guangdong-Hong Kong Joint Laboratory of Quantum Matter, Department of Physics, \\ and HK Institute of Quantum Science \& Technology,\\ The University of Hong Kong, Pokfulam Road, Hong Kong, China}
	
        \author{Jianwen Xu}\thanks{These authors contributed equally to this work.}
	\affiliation{National Laboratory of Solid State Microstructures, School of Physics,
		Nanjing University, Nanjing 210093, China}
		\affiliation{Shishan Laboratory, Suzhou Campus of Nanjing University, Suzhou 215000, China}
  
	\author{Wen Zheng}
     \email{zhengwen@nju.edu.cn}
	\affiliation{National Laboratory of Solid State Microstructures, School of Physics,
		Nanjing University, Nanjing 210093, China}
	
		\affiliation{Shishan Laboratory, Suzhou Campus of Nanjing University, Suzhou 215000, China}

       \author{Dong Lan}
	\affiliation{National Laboratory of Solid State Microstructures, School of Physics, Nanjing University, Nanjing 210093, China}
		\affiliation{Shishan Laboratory, Suzhou Campus of Nanjing University, Suzhou 215000, China}
     \affiliation{Hefei National Laboratory, Hefei 230088, China}

	\author{Giandomenico Palumbo}
	\affiliation{School of Theoretical Physics, Dublin Institute for Advanced Studies, 10 Burlington Road, Dublin 4, Ireland}
	
	\author{Nathan Goldman}
	\affiliation{Center for Nonlinear Phenomena and Complex Systems,
		Universit\'{e} Libre de Bruxelles, CP 231, Campus Plaine, B-1050 Brussels, Belgium}
	
	\author{Shi-Liang Zhu}
	\affiliation{Guangdong-Hong Kong Joint Laboratory of Quantum Matter,
		Frontier Research Institute for Physics, South China Normal University, Guangzhou 510006, China}
	\affiliation{Guangdong Provincial Key Laboratory of Quantum Engineering and Quantum Materials,
		School of Physics and Telecommunication Engineering,
		South China Normal University, Guangzhou 510006, China}

	\author{Xinsheng Tan}
	\email{tanxs@nju.edu.cn}
	\affiliation{National Laboratory of Solid State Microstructures, School of Physics,
		Nanjing University, Nanjing 210093, China}
		\affiliation{Shishan Laboratory, Suzhou Campus of Nanjing University, Suzhou 215000, China}
	\affiliation{Hefei National Laboratory, Hefei 230088, China}
 
    \author{Z. D. Wang}
	  \email{zwang@hku.hk}
	\affiliation{Guangdong-Hong Kong Joint Laboratory of Quantum Matter, Department of Physics, \\ and HK Institute of Quantum Science \& Technology,\\ The University of Hong Kong, Pokfulam Road, Hong Kong, China}
	
	\author{Yang Yu}
	\affiliation{National Laboratory of Solid State Microstructures, School of Physics,
		Nanjing University, Nanjing 210093, China}
		\affiliation{Shishan Laboratory, Suzhou Campus of Nanjing University, Suzhou 215000, China}
    \affiliation{Hefei National Laboratory, Hefei 230088, China}

	\begin{abstract}    
		{We present the successful realization of four-dimensional (4D) semimetal bands featuring tensor monopoles, achieved using superconducting quantum circuits. Our experiment involves the creation of a highly tunable diamond energy diagram with four coupled transmons, and the parametric modulation of their tunable couplers, effectively mapping momentum space to parameter space. This approach enables us to establish a 4D Dirac-like Hamiltonian with fourfold degenerate points. Moreover, we manipulate the energy of tensor monopoles by introducing an additional pump microwave field, generating effective magnetic and pseudo-electric fields and simulating topological parity magnetic effects emerging from the parity anomaly. Utilizing non-adiabatic response methods, we measure the fractional second Chern number for a Dirac valley with a varying mass term, signifying a nontrivial topological phase transition connected to a 5D Yang monopole. Our work lays the foundation for further investigations into higher-dimensional topological states of matter and enriches our comprehension of topological phenomena.

		}    
	\end{abstract}

	\maketitle

	\emph{Introduction.---}
    Topological phases of matter have attracted broad interest from different physical communities ranging from condensed matter\cite{Hasan2010,Qi2011,Armitage2018} to synthetic systems\cite{Zhang2018,Cooper2019,YXu2019,Ozawa2019,WZhu2023}, being at the heart of modern physics. One of the intriguing and exciting features is that these novel phases support topological electromagnetic responses described by the associated topological field theory \cite{RABertlmann2000, fradkin_2013}.
    For instance, in the well-known 3D Weyl semimetals, the chiral magnetic effect is induced by a paired Weyl dipole resulting from the chiral anomaly \cite{AAZyuzin2012,MMVazifeh2013,DIPikulin2016,JBehrends2019,ZZheng2019}.  In 2D dipolar Dirac semimetals, parity anomaly induces the pseudo-Hall effect \cite{ZLin2019} with the pseudo-electric field arising from an external strain field\cite{Shapourian2015,Cortijo2015,Grushin2016}. Additionally, significant attention has also been given to the quantum anomalous Hall effect in 2D Chern insulators and the topological magnetoelectric effect in 3D topological insulators\cite{XLQi2008}. 
		
By contrast, topological phases and the related electromagnetic effects in higher dimensions are much less studied, especially in experiments. This is primarily due to the constraint of natural materials that can be used in such experiments.  Fortunately, quantum simulations using synthetic matter have emerged as a powerful tool for simulating higher dimensional topological states of matter, owing to their high flexibility and controllability. For instance, the 4D quantum Hall physics \cite{SCZhang2001,Karabali2002,XLQi2008,Price2015} and the corresponding 
Thouless pumping in 2D \cite{Kraus2013} has been experimentally realized in ultracold atoms \cite{Lohse2018,Bouhiron2022} and photonics \cite{Zilberberg2018}. Recently, a distinct kind of 4D tensor monopole, characterized by the first Dixmier-Douady (DD) invariant, has been theoretically investigated \cite{Palumbo2018,Palumbo2019,YQZhu2020,Weisbrich2021tensormonopolesin,HTDing2020} and experimentally realized in superconducting circuits \cite{XSTan2021} and NV centers \cite{MoChen2022}.  In addition, the parity magnetic effect arising from the parity anomaly in (4+1)D induced by a pair of 4D monopoles has been predicted in a 4D topological semimetal \cite{YQZhu2020}. However, no experimental demonstration is reported about this novel topological electromagnetic effect so far.

In this Letter, we emulate a 4D topological semimetal band with tensor monopoles by using superconducting quantum circuits. We show how to detect the parity magnetic effect in this system by generating an effective magnetic (pseudo-electric) field and measuring the second Chern number (SCN) of a massive Dirac-like cone after introducing a mass regulator for a 4D monopole. Interestingly, the corresponding gapped system with two Dirac-like valleys hosts opposite fractional SCN, which exhibits a so-called valley-induced magnetic effect similar to the gapless case. Meanwhile, this gapped system can be characterized by a second valley Chern number defined as a half difference of the SCN for two valleys, and its behavior follows the 4D valley Hall effect predicted in the recent work \cite{YQZhu2022}. Similar methods in our superconducting system can also detect this second valley Chern number and valley Hall effect. We also present the discussion about the topological phase transition based on this 4D model.

	\begin{figure}
		\centering
		\includegraphics[]{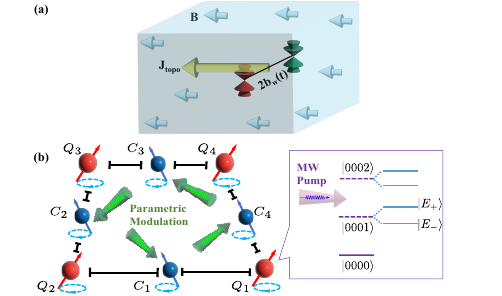}
		\caption{ 
			(a) Schematic of parity magnetic effect. A pair of 4D monopoles exist in the momentum space. The separation $b_w(t)$ is modulated artificially to create an effective electric field $E^w_5=\partial_t b_w$. In the presence of a static magnetic field $\mathbf{B}$, a topological current $J_{topo}$ will be induced.
			(b) The device consists of four transmons(red) and four couplers(blue). Four parametric modulations with different frequencies generate the Hamiltonian(green arrows), while an additional   pump signal is applied to $Q_1$ to simulate the fictitious chemical potential. The lower energy level $|E_-\rangle$ and the first excited states of other qubits constitute the adjustable first excited manifold. \label{fig:FiG1}}
	\end{figure}

	\emph{Experimental setup.--}The layout of our device is shown in Fig. \ref{fig:FiG1}, where four transmon qubits form a $2\times2$ square lattice \cite{JKoch2007, RBarends2013, JQYou2011, Arute2019, YLWu2021}. Each side of the square consists of two transmons connected to a center tunable coupler with the same coupling strength and further coupled to each other directly through a residue capacitance. Each qubit and coupler is labeled as $Q_i$ and $C_j$ respectively, where $ i,j\in\{1,2,3,4\}$.  All qubits are designed negatively detuned from the nearest coupler, $\Delta_{i j}=\omega_{Q_i}-\omega_{C_j}<0$. The system quantum crosstalk has been analyzed recently in this kind of multi-qubit lattice \cite{PZhao2022, DMZajac2021spectator, JChu2021}. In order to adiabatically eliminate the couplers, the system operates in the dispersive regime, i.e., $g\ll|\Delta|$. The entire Hamiltonian of the system can be written as (the coupling of the below Hamiltonian only exists in the nearest neighbor):
	\begin{equation}
		\begin{split}
			H/{\hbar}&=\sum_{i,j=1}^4 \left[-\frac{1}{2}\omega_{Q_i}\sigma_{Q_i}^z - \frac{1}{2}\omega_{C_j}(\phi_j)\sigma_{C_j}^z\right. \\ 
			&\left. +(g_{i, j}\sigma_{Q_i}^+ \sigma_{C_j}^- +  g_{i+1, j}\sigma_{Q_{i+1}}^+ \sigma_{C_j}^- + \text{H.c.})\right]
		\end{split}
	\end{equation}

	\begin{figure*}[t]
		\centering
		\includegraphics[]{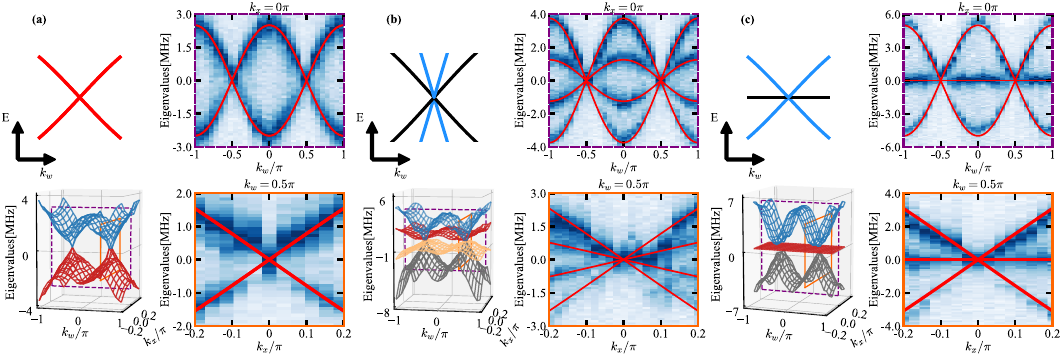}  
		\caption{ Different spectrum $E(k_w)$ at FBZ for $a=0$, $0.5$, and $1$ are shown in (a), (b), and (c), respectively. The upper left panels are the schematic energy dispersion $E(k_w)$ at $k_{x,y,z}=0$ near the right nodal point. The lower left panels demonstrate the measured energy spectrum of 4D topological semimetals by changing $k_w$ and $k_x$. The purple dash frame and the orange solid frame mark the cross-sections corresponding to the data in the right panels. The 1D spectra are shown in the upper and lower right panels. The theoretical calculation (red lines) agrees well with the experimental results. \label{fig:FiG2}}
	\end{figure*}

The approach of parametric modulation provides more degrees of freedom in parameters, making it convenient to realize the designed Hamiltonian.  The effective coupling strength between the adjacent qubits depends on the frequency of the coupler $J_{i,i+1}=g_{i,i+1}+g_{i,j}g_{i+1,j}(1/\Delta_{i,j}+1/\Delta_{i+1,j})/2$ \cite{FYan2018, EASete2021, JStehlik2021}, we apply four independent sinusoidal fast-flux biases $\phi_j(t)=\Phi_j + \delta_j \cos(\omega_{\phi_j}t+\varphi_j)$ to coupler $C_j$ respectively, and obtain resonant exchange interaction $\Omega_{i,i+1}=\frac{\delta_j}{2} \frac{\partial J_{i,i+1}}{\partial \phi_j}e^{-i\varphi_j}$ \cite{DCMcKay2016, MRoth2017, MReagor2018, MGanzhorn2019, PMundada2019, MGanzhorn2020}. 
	
	In the first-excitation subspace, we exchange the order of $Q_1$ and $Q_2$, the Hamiltonian can be written as
	\begin{equation}
		\begin{aligned}
			H_{\text{eff}}/\hbar = 
			\begin{bmatrix}
				0 & \Omega_{12}^* & \Omega_{23} & 0 \\
				\Omega_{12} & 0 & 0 & \Omega_{41}^* \\
				\Omega_{23}^* & 0 & 0 & \Omega_{34} \\
				0 & \Omega_{41} & \Omega_{34}^* & 0
			\end{bmatrix},
		\end{aligned}
	\end{equation}
showing a diamond configuration. Here the coupling strengths and phases are tunable, and both can be easily calibrated \cite{SM}. 

\emph{Topological model and response.---} By mapping the system parameters into the Bloch vectors, i.e., $\Omega_{12}=d_x+ad_w+i(d_z+ad_y)$,$\Omega_{23}=d_w+ad_x-i(d_y+ad_z)$,$\Omega_{34}=d_x-ad_w+i(-d_z+ad_y)$, and $\Omega_{41}=-d_w+ad_x+i(-d_y+ad_z)$, we demonstrate that our diamond-type coupling can be used to simulate the momentum-space Hamiltonian \cite{YQZhu2020} as 
	\begin{equation}
		\begin{aligned}
			H({\boldsymbol k})=d_x \tilde{\Gamma}_x + d_y \tilde{\Gamma}_y + d_z \tilde{\Gamma}_z + d_w \tilde{\Gamma}_w
		\end{aligned}
		\label{Ham_mo}
	\end{equation}
	with the four-component Bloch vector
	$d_i=v_i \sin k_i$ for $i\in\{x,y,z\}$, and $d_w=v_w(\Lambda+3-\sum_{j=x,y,z,w}\cos k_j)$. Here $v_i$ denotes the Fermi velocity, and $\Lambda$ is a tunable parameter.  The matrices $\tilde{\Gamma}_x = \sigma_0 \otimes \sigma_1 + a\sigma_1 \otimes \sigma_0, \tilde{\Gamma}_y = \sigma_2 \otimes \sigma_3 + a\sigma_3 \otimes \sigma_2,\tilde{\Gamma}_z = \sigma_0 \otimes \sigma_2 + a\sigma_2 \otimes \sigma_0, \tilde{\Gamma}_w = \sigma_1 \otimes \sigma_3 + a\sigma_3 \otimes \sigma_1$,
	where $a$ is constant and $\sigma_k$ are the standard Pauli matrices, and $\sigma_0$ being the identity matrix.  
	
    Note that this model always preserves chiral symmetry, i.e., $\{\mathcal{S}, H\}=0$ with $\mathcal{S}=\tilde{\Gamma}_0=\sigma_3\otimes\sigma_3$. The energy spectrum of this model is given by $E=\pm(1\pm a)\sqrt{\sum_jd_j^2}$. For $\Lambda<|1|$, the system hosts two monopoles near $\boldsymbol K_{\pm}=(0,0,0,\pm \arccos \Lambda)$ with the effective Hamiltonian
    \begin{equation}
    H_{\pm}({\boldsymbol q}^{\pm})=q^{\pm}_x\tilde{\Gamma}_x+q^{\pm}_y\tilde{\Gamma}_y+q^{\pm}_z\tilde{\Gamma}_z\pm\beta q^{\pm}_w\tilde{\Gamma}_w,
    \end{equation}
where $\beta=\sqrt{1-\Lambda^2}$, and ${\boldsymbol q}^{\pm}={\boldsymbol k}-\boldsymbol K_{\pm}$. There are two types of monopoles for $H_{\pm}$ when $a$ takes different values. For $a=0$, it is a $\mathbb{Z}_2$ monopole characterized by a 3D winding number $w=1$ protected by an extra $CP$ symmetry, i.e., $\{\mathcal{CP},H\}=0$ with $\mathcal{CP}=\sigma_1\otimes\sigma_2\mathcal{K}$ satisfying $(\mathcal{CP})^2=-1$. For $a\neq 0$, $H_{\pm}$ breaks $CP$ but only keeps chiral symmetries, which is coined tensor monopoles characterized by the Dixmier-Douady ($\mathcal{DD}$) invariant belong to $\mathbb{Z}$ class\cite{Palumbo2018,Palumbo2019}. Notably, a tensor monopole $H_{\pm}$ hosts the topological charge $\mathcal{DD}=\pm 2$ for $a\neq 0,\pm 1$ while $\mathcal{DD}=\pm 1$ for $a=\pm 1$ due to the flat limit. All the above topological numbers are defined on the 3D hypersphere enclosing the monopole. Thus it implies this system presents a monopole-to-monopole topological phase transition by tuning the parameter $a$ \cite{YQZhu2020}. 
 
To experimentally visualize the band configuration of this 4D topological semimetal, we measure the spectra of the Hamiltonian $H({\boldsymbol k})$ in the first Brillouin zone(FBZ). 
As shown in Fig. \ref{fig:FiG2}(a), we measure two two-fold degenerate energy bands with a pair of $\mathbb{Z}_2$ monopoles for the spin-1/2 case when $a=0$. As we enlarge parameter $a$ to be $0<|a|<1$, the degeneracy for each monopole is lifted, which gives a tensor semimetal phase with four energy bands as shown in Fig. \ref{fig:FiG2}(b) with $a=0.5$. Especially when $a=\pm1$, two middle bands become perfectly flat, so we can only observe three energy levels near the tensor monopoles, as illustrated in Fig. \ref{fig:FiG2}(c). 
 In Fig. \ref{fig:FiG2}(a-c), we set $\Lambda=0$ for simplicity. Additionally, we can vary the parameter $\Lambda$ without difficulty tuning the separation between two monopoles.
	
	\begin{figure*}
		\centering
		\includegraphics[]{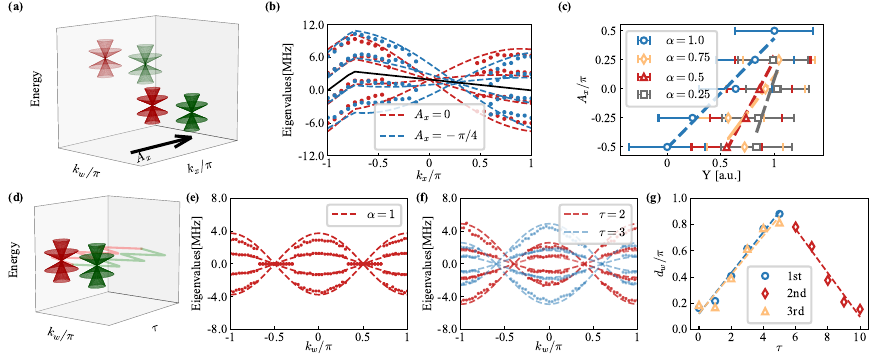}
		\caption{Experiment results for $a=0.5$ (a) Diagrams of the effects of $A_x$ used to apply effective magnetic field. (b) Spectrum along the $k_x$ direction for $u_0(\textbf{k})=f(k_x)$   (black line). The vector potential $A_x$ is set to 0 and $-\pi/4$, respectively. The dashed lines are the theoretical results, and the dots are the extracted experiment data. (c) $A_x$ varies with $Y$ at different $\alpha=1.0$, $0.75$, $0.5$ and $0.25$, extracting different magnetic field. (d) Illustration of the modulation of the monopole separation under a magnetic field. (e) Spectrum along the $k_w$ axis without modulation ($\Lambda=0$) under a static magnetic field. (f) Spectrum at different discrete time $\tau$ under a time dependent modulation $\Lambda(\tau)$ for $\tau=2$ (red) and $\tau=3$ (blue) with $\alpha=1.0$. (g) The monopole separation changes in different magnetic fields with fictitious time. For $\alpha=1.0$, $\tau$ changes from 0 to 5 (blue) and 6 to 10 (red), and for $\alpha=0.5$, $\tau$ changes from 0 to 5 (yellow).}
		\label{fig:FiG3}
	\end{figure*}	

For such a 4D semimetal phase with a pair of monopoles with opposite charges separated along $k_w$-direction with the distance $|{\boldsymbol{K}}_+-{\boldsymbol{K}}_-|=2b_w$, an anomalous topological current will be generated in the presence of a magnetic field $\mathbf{B}=B^z\hat{e}_z$ and and a pseudo-electric field $E_5^w=\partial_t b_w$\cite{Cortijo2015, DIPikulin2016, Grushin2016, Roy_2018}, which is given by\cite{YQZhu2020}, 
	\begin{equation}
		\begin{aligned}
		J^{z}=\frac{C_2^+}{2{\pi}^2}E_5^wB^z.
		\end{aligned}\label{res_cur}
	\end{equation}
Note that $C_2^+$ is the SCN for a single Dirac-like valley near ${\boldsymbol K}_+$ described by $H_{v,+}=H_++\tilde{m}$ when performing the Pauli-Villars method by introducing a chiral-broken mass regulator $\tilde{m}=m \tilde{\Gamma}_0$ to open the gap. $C_2^+=\text{sgn}(m)/2$ for $a\neq \pm 1$ while $C_2^+=\text{sgn}(m)/8$ for $a=\pm 1$ \cite{SM}. Namely, the topological current induced by two tensor monopoles in the flat limit is only $1/4$ of those cases for $a\neq \pm 1$. Similarly, another valley near ${\boldsymbol K}_-$ described by $H_{v,-}=H_{-}+\tilde{m}$ hosts an opposite SCN, i.e., $C_{2,-}=-C_{2,+}$, contrasting to $H_{v,+}$. PME in the semimetal phase ($m=0$) without gap shares the same results as its corresponding gaped system ($m\neq 0$) when we effectively introduce a chiral-broken mass term $\tilde{m}$ into $H({\boldsymbol k})$. The topological response is also named ``valley-induced magnetic effect," which has been well-studied in Ref. \cite{YQZhu2022} recently. The total system $H_{\text{tot}}=H+\tilde{m}$ now can be characterized by the second valley Chern number $C_{2,v}=(C_2^+-C_2^-)/2=C_2^+$. Therefore, this system also supports the so-call ``4D quantum valley Hall effect" with valley current as $J^z_5=\frac{C_{2v}}{4\pi^2}E^zB^z$ , where two valleys contribute opposite Hall currents \cite{YQZhu2022}.

In the following, we show how to emulate this PME in our superconducting system. In order to generate a magnetic field, we introduce a momentum-dependent energy shift $u_0(\boldsymbol{k})$, $H_{\text{pump}}/\hbar = H_{\text{eff}}/\hbar + u_0(\textbf{k})I_4$.
We utilize the Autler-Townes splitting (ATS) to manipulate the energy shift \cite{BRMollow1969, MBaur2009, Sillanpaa2009prl, HCSun2014pra, XSTan2019}. Specifically, we apply an extra pump microwave field $\Omega_d \cos(\omega_d  t + \varphi_d)(\sigma_{Q_1}^+ + \sigma_{Q_1}^-)$ to $Q_1$ with frequency $\omega_d = \omega_{12}^{Q_1}$.  The first excited state splits into $|E_-\rangle$ and $|E_+\rangle$, and the splitting separation is controlled by the Rabi frequency $\Omega_d$ \cite{SM}. The lower spectrum $|E_-\rangle$ is treated as a new controllable excited state with modified parametric modulation.  By carefully designing the parameters,  we alleviate the disturbance of $|E_+\rangle$  to our four-band Hamiltonian. 
	
	We design the $\boldsymbol k$-dependent $u_0$ to be constant with different $k_{y,z,w}$, $u_0(\boldsymbol k)=f(k_x)$. $u_{\max}/2\pi=3.46$ MHz $f(k_x)=\alpha (\frac{4u_{\max}}{\pi}) (k_x+\pi) $ for $-\pi\leq k_x \leq -3\pi/4$, whereas $f(k_x)=-\alpha (\frac{4u_{\max}}{7\pi})(k_x - \pi)$  for  $-3\pi/4 \leq k_x\leq \pi$. In the range $[-3\pi/4, \pi]$, the Rabi frequency of the pump microwave field changes linearly. Fig. \ref{fig:FiG3}(a) demonstrates that two monopoles can move along $k_x$ axis in the same direction with the change of $A_x$, while the separation remains unchanged. We set a series of different $A_x$, then measure the energy shift $\Delta E$ of one monopole. This shift can be considered as the response of a fictitious force $F_Y=1$, $\Delta E=F_Y\Delta Y$ \cite{XSTan2019}. As indicated by Fig. \ref{fig:FiG3}(b), take $a=0.5$ for example, we set $A_x=\pi/4$ and extract the eigen-energies from the spectrum data (blue), the shift of monopoles is obvious compared to $A_x=0$ (red). For $\alpha=1.0$, we set $A_x \in \{-2\pi/4,-\pi/4,0,\pi/4,2\pi/4\}$ and find expected good linear relation between $A_x$ and $Y$, the slope gives the magnetic field $B^z=-\partial_YA_x$. We can increase the magnetic field by changing $\alpha$ from 1.0 to 0.25, as shown in Fig. \ref{fig:FiG3}(c).

	\emph{Pseudo-electric field.---} It is extremely difficult to  observe the topological current directly in superconducting circuits. In practice, we set the separation as a function of a fictitious time $\tau$ such that the partial derivative of the separation concerning $\tau$ is a constant. We set $\Lambda(\tau)=\cos(0.2\pi\times \tau)$, then $ \partial _\tau b_w = 0.2\pi$. The monopole separation is obtained from the spectrum measured with a fixed $\Lambda$ in a discrete-time $\tau$. As shown in Fig. \ref{fig:FiG3}(f), with the time changing from $2$ to $3$ for $\alpha=1.0$, the separation between the two monopoles varies from $0.4\pi$ to $0.6\pi$. At the same time, we can set different $\alpha$ and provide different magnetic fields while modulating the monopole separation. In the first fictitious modulation, two monopoles start from $k_w=0$ and move in the opposite direction. And in the second modulation, two monopoles start from the two boundaries of the FBZ, move closer, and merge. The third modulation is almost the same as the first except $\alpha=0.5$. We show the pseudo-electric field $E^w_5=\partial_t b_w$ in Fig. \ref{fig:FiG3}(g).
	
	\emph{Measuring Second Chern number.---} In this section, we show how to measure the SCN for $H_{v,+}$. 
	For simplicity, we consider spin-1/2 case when $a=0$ with $C_2=\text{sgn}(m)/2$  and label ${\boldsymbol q}^+$ as ${\boldsymbol q}$ hereafter. By parametrizing ${\boldsymbol q}$ into the Hopf coordinates, we obtain: $q_x=q\cos\theta\cos\phi$, $q_y=q\cos\theta\sin\phi$, $q_z=q\sin\theta\cos\varphi$, and $q_w=q\sin\theta \sin\varphi$. This Dirac valley consists of two bands $E_{\pm}=\pm \sqrt{q^2+m^2}$, and each band has two-fold degeneracy.
	In particular, we denote the lower occupied degenerate bands with eigenvalues $E_n({\boldsymbol q})=E_-$ and eigenstates $|u_{n}({\boldsymbol q})\rangle$ where $n \in \{1,2\}$.
	At this time, the SCN can be calculated in the Hopf coordinates as,
	\begin{equation}
		\begin{aligned}
			C_2= &\frac{1}{4\pi^2}\int_{\mathbb{R}^4} \text {tr}(\mathcal{F}_{q\theta}\mathcal{F}_{\phi \varphi}+\mathcal{F}_{\varphi q}\mathcal{F}_{\phi \theta}+\mathcal{F}_{\phi q}\mathcal{F}_{\theta \varphi}) d^4 q \\
&=  \int_{0}^{\infty} dq \int_{0}^{\pi/2} \mathcal{FF}(q,\theta) d\theta  \int_{0}^{2\pi}\int_{0}^{2\pi} d\phi d\varphi,
		\end{aligned}
	\end{equation}
where the non-Abelian Berry curvature is defined as
$\mathcal{F}_{\mu\nu}=\partial_{\mu}\mathcal{A}_{\nu}-\partial_{\nu}\mathcal{A}_{\mu}-i[\mathcal{A}_{\mu}, \mathcal{A}_{\nu}]$ with the associated non-Abelian Berry connection $\mathcal{A}_{\mu}^{\alpha\beta} = i\langle u_{\alpha} |\partial_{\mu}|u_{\beta}\rangle$. Using the analytical relations $\text{tr}(\mathcal{F}_{q\theta}\mathcal{F}_{\phi \varphi})= \text{tr}(\mathcal{F}_{\varphi q}\mathcal{F}_{\phi \theta})=\text{tr}(\mathcal{F}_{\phi q}\mathcal{F}_{\theta \varphi})$, we obtain $\mathcal{FF}(q,\theta)=\frac{3mq^3\cos\theta\sin\theta}{8\pi^2(m^2+q^2)^{5/2}}$. Notice that $\mathcal{FF}(q,\theta)$ is  independent to $\phi$ and $\varphi$,  without loss of generality, we set $(\phi,\varphi)=(0,0)$ in the measurement. The Berry curvature components  $\mathcal{F}_{q\theta}$ and $\mathcal{F}_{\phi \varphi}$ can be measured through the generalized geometric force in $(q,\theta)$ space\cite{SM}. Therefore, during measurement, these quantities evolve into the tomography of each qubit with varying parameters in the 2D subspace $(q,\theta)$. To ensure the feasibility, we set a cutoff for spatial region $q$, i.e., $q\in[0,q_{\text{cut}}]$.
	\begin{figure}
		\centering
		\includegraphics[]{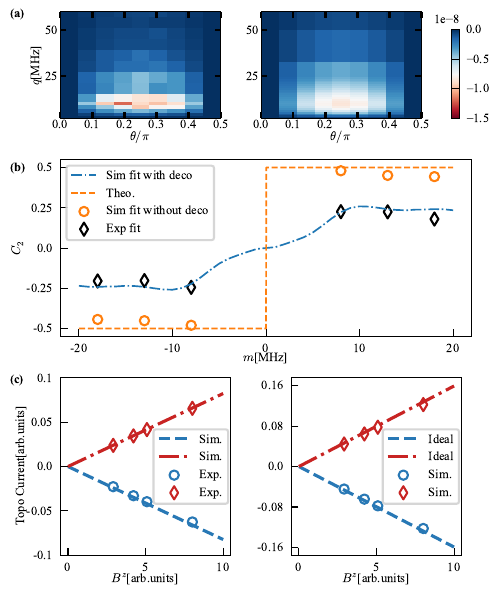}
		\caption{SCN $C_2$, Chern form $\mathcal{FF}/4\pi^2$, and current $J^z$. (a) Experimental (left) and simulation (right) results of Chern form  $\mathcal{FF}(q,\theta)/4\pi^2$ at different $q$ and $\theta$. (b) The second Chern number $C_2$ for different $m$. Black dots represent the experiment result, which agrees with the numerical simulation result after considering the decoherence, as shown by the blue dashed line. The orange dashed line is the theoretical result $C_2=\text{sgn}(m)/2$. Orange dots show the numerical simulation without the decoherence. 
          (c) Left panel: For different magnetic fields and modulations, experimental data and simulation results with decoherence of topological current are plotted by dots and dash lines. Right panel: Simulation results (dots) without considering the decoherence align with the ideal results (lines).
              }
		\label{fig:FiG4}
	\end{figure}
	In this experiment, we measure the first-order non-adiabatic response and get the components of non-Abelian Berry curvature at $\theta \in [0, \pi/2]$ and $q\in[0,q_{\text{cut}}]$, where $q_{\text{cut}}=200$ MHz by slowly ramping corresponding parameters \cite{VGritsev2012, MDSchroer2014, roushan2014observation, MKolodrubetz2016, SSugawa2018}. We use unitary transformation to decouple the system into two two-level systems and to increase the signal-to-noise ratio, and we apply modified TQDA protocol \cite{SM, Berry2009, demirplak2008consistency, Campo2013prl, UNANYAN1997, XChen2010prl}. In Fig. \ref{fig:FiG4}(a), we illustrate experimental data and numerical simulation \cite{JOHANSSON2012, JOHANSSON2013} of Berry curvature for $m=8$ MHz. The measurement results of SCN are shown in Fig. \ref{fig:FiG4}(b). It will agree with the theoretical result with improved qubit coherence. 
	In addition, with the constructed fictitious magnetic field and time-varying gapless points, we can measure the topological currents according to Eq.~\eqref{res_cur}. Fig. \ref{fig:FiG4}(c) shows the good linear relationship between topological current $J^z$ and fictitious magnetic field $B^z$ we simulated. Compared with the theoretical prediction, the slope is slightly smaller, mainly caused by the short decoherence time. Ideal results can be obtained by this approach as shown in Fig.~\ref{fig:FiG4}(d).

\emph{Conclusion and outlook.---}
In summary, we present the first experimental report of the quantum emulation of the PME with 4D monopoles using superconducting
qubits.  Additionally, we measure the fractional SCN for a 4D Dirac valley with varying mass $m$, indicating a topological phase transition for a 4D Dirac valley ($a=0$) from $C_2=-1/2$ to $C_2=1/2$ for a 4D Dirac valley. By treating $m$ as the fifth dimension, it represents a 5D Yang monopole \cite{Yang1978,SSugawa2018} that hosts the topological charge as $Q=\Delta C_2=1$.  Notably, when $a=\pm1$, the corresponding 5D defect \cite{SM} gives rise to a new type of Nexus quadrupole points.  This higher-dimensional generalization of the 3D Nexus triple points \cite{Das2023} presents an intriguing avenue for future exploration.
 Our work  opens the possibility to explore more complex topological systems in higher dimensions using a highly tunable platform.

\begin{acknowledgements}
This work was partly supported by the Key R\&D Program of Guangdong Province (Grant No. 2018B030326001), NSFC (Grant No. 12074179, No. 11890704, and No. U21A20436), NSF of Jiangsu Province (Grant No. BE2021015-1), NSFC/RGC JRS grant (N-HKU774/21), the CRF of Hong Kong (C6009-20G), and Innovation Program for Quantum Science and Technology (2021ZD0301700).
\end{acknowledgements}

\end{document}